\documentclass[lettersize,journal]{IEEEtran}
\usepackage{amsmath,amsfonts}
\usepackage{algorithm}
\usepackage{algorithmicx}
\usepackage{algpseudocode}
\usepackage{multirow}
\usepackage{booktabs}
\usepackage{array}
\usepackage[caption=false,font=footnotesize,labelfont=rm,textfont=rm]{subfig}
\usepackage{textcomp}
\usepackage{stfloats}
\usepackage{url}
\usepackage{verbatim}
 \usepackage{cite}
\usepackage{graphicx}
\usepackage{float}                       
\usepackage{overpic} 
\usepackage{diagbox}
\def\BibTeX{{\rm B\kern-.05em{\sc i\kern-.025em b}\kern-.08em
    T\kern-.1667em\lower.7ex\hbox{E}\kern-.125emX}}
\usepackage{balance}
\begin{document}
\title{Spiking Semantic Communication for Feature Transmission with HARQ}
\author{Mengyang Wang, Jiahui Li, \IEEEmembership{Member, IEEE}, Mengyao Ma, \IEEEmembership{Member, IEEE}, \\ Xiaopeng Fan, \IEEEmembership{Senior Member, IEEE}
\thanks{This work was supported in part by the National Key R\&D Program of China (2021YFF0900500), the National Natural Science Foundation of China (NSFC) under grants U22B2035 and 62272128.  \textit{(Corresponding author: Xiaopeng Fan, Mengyao Ma.)}

Mengyang Wang and Xiaopeng Fan are with the School of Computer Science, Harbin Institute of Technology, Harbin 150001, China, and also with Peng Cheng Laboratory, Shenzhen 519055, China (e-mail: mywang1996@outlook.com; fxp@hit.edu.cn). 

Jiahui Li and Mengyao Ma are with the Wireless Technology Lab, Huawei, Shenzhen 518129, China (e-mail: lijiahui666@huawei.com; ma.mengyao@huawei.com). }}

\maketitle

\begin{abstract}
In Collaborative Intelligence (CI), the Artificial Intelligence (AI) model is divided between the edge and the cloud, with intermediate features being sent from the edge to the cloud for inference. Several deep learning-based Semantic Communication (SC) models have been proposed to reduce feature transmission overhead and mitigate channel noise interference. Previous research has demonstrated that Spiking Neural Network (SNN)-based SC models exhibit greater robustness on digital channels compared to Deep Neural Network (DNN)-based SC models. However, the existing SNN-based SC models require fixed time steps, resulting in fixed transmission bandwidths that cannot be adaptively adjusted based on channel conditions. To address this issue, this paper introduces a novel SC model called SNN-SC-HARQ, which combines the SNN-based SC model with the Hybrid Automatic Repeat Request (HARQ) mechanism. SNN-SC-HARQ comprises an SNN-based SC model that supports the transmission of features at varying bandwidths, along with a policy model that determines the appropriate bandwidth. Experimental results show that SNN-SC-HARQ can dynamically adjust the bandwidth according to the channel conditions without performance loss.
\end{abstract}


\section{Introduction}
\IEEEPARstart{D}{eep} Neural Network (DNN)-based models have demonstrated effectiveness across various tasks. However, their deployment on edge devices is hindered by the substantial demand for computing resources. To address this issue, Collaborative Intelligence (CI) \cite{2} has emerged as a promising solution. CI splits a neural network between an edge device and a cloud server, allowing for task inference. Nevertheless, the transmission of the original feature from the edge device to the cloud server incurs significant overhead, compounded by the unreliability of wireless channels. In this paper, Semantic Communication (SC) is used to tackle these problems.

In contrast to the Shannon paradigm, SC only transmits relevant semantic information for a specific task to the receiver, resulting in a significant reduction in data traffic \cite{28}. Recently, many Deep Learning (DL)-based SC models have been developed to transmit features \cite{3,4,5,6,7}. DNN-based SC models in \cite{3} and \cite{5} extract compact semantic information from features and transmit it through the AWGN channel. To ensure compatibility with digital communication systems, the extracted semantic information must be converted into bits for transmission. In \cite{4}, additional quantization is employed to binarize the output of the DNN-based SC model. On the other hand, SNN-based SC models in \cite{6} and \cite{7} directly transmit semantic information on the digital channel due to the binary output of the SNN. Furthermore, both studies demonstrate that the SNN-based SC model exhibits greater robustness than the DNN-based SC model under poor channel conditions.

However, the SNNs in \cite{6} and \cite{7} are time-step-driven, requiring multiple time steps to transmit the feature in the SC model. The time step is manually set, resulting in a fixed amount of transmitted data (i.e., bandwidth). To address this limitation, it is desirable to build an SNN-based SC model that can dynamically adjust the running time step, thereby adjusting the transmission bandwidth, while maintaining satisfactory performance for cloud tasks. In this paper, we propose a novel SC model called SNN-SC-HARQ, which efficiently transmits features by incorporating the mechanism of Hybrid Automatic Repeat Request (HARQ) into our previous work, SNN-SC \cite{6}. Previous studies on DNN-based SC for text transmission have considered HARQ \cite{10,11}, but they require multiple pairs of codec models to support different bandwidths, which is storage-consuming. In contrast, SNN-SC-HARQ only needs to adjust the running time step to change transmission bandwidths, making it efficient and saving storage space. Our contributions are summarized as follows:
\begin{itemize}[\topsep=0pt]
\item{We propose multi-rate SNN-SC, which can transmit features at varying bandwidths using a single model.}
\item{We propose a new metric called semantic similarity, which measures the similarity between reconstructed and original features at the semantic level. To estimate semantic similarity, we propose SimNet. By comparing the estimated semantic similarity with a preset threshold, SimNet is capable of making decisions at receiver.}
\item{By combining multi-rate SNN-SC with SimNet, we obtain SNN-SC-HARQ. This model is superior to the manual bandwidth adjustment model as it can select an appropriate bandwidth based on channel conditions. By setting different thresholds in SimNet, we can achieve a balance between bandwidth and task performance.}
\end{itemize}

The rest part of this paper is as follows. Section \ref{sec:2} details the proposed SNN-SC-HARQ. Numerical results are presented in Section \ref{sec:3}. Finally, the paper is concluded in Section \ref{sec:4}.
\begin{figure*}
  \centering
  \includegraphics[width=1.75\columnwidth]{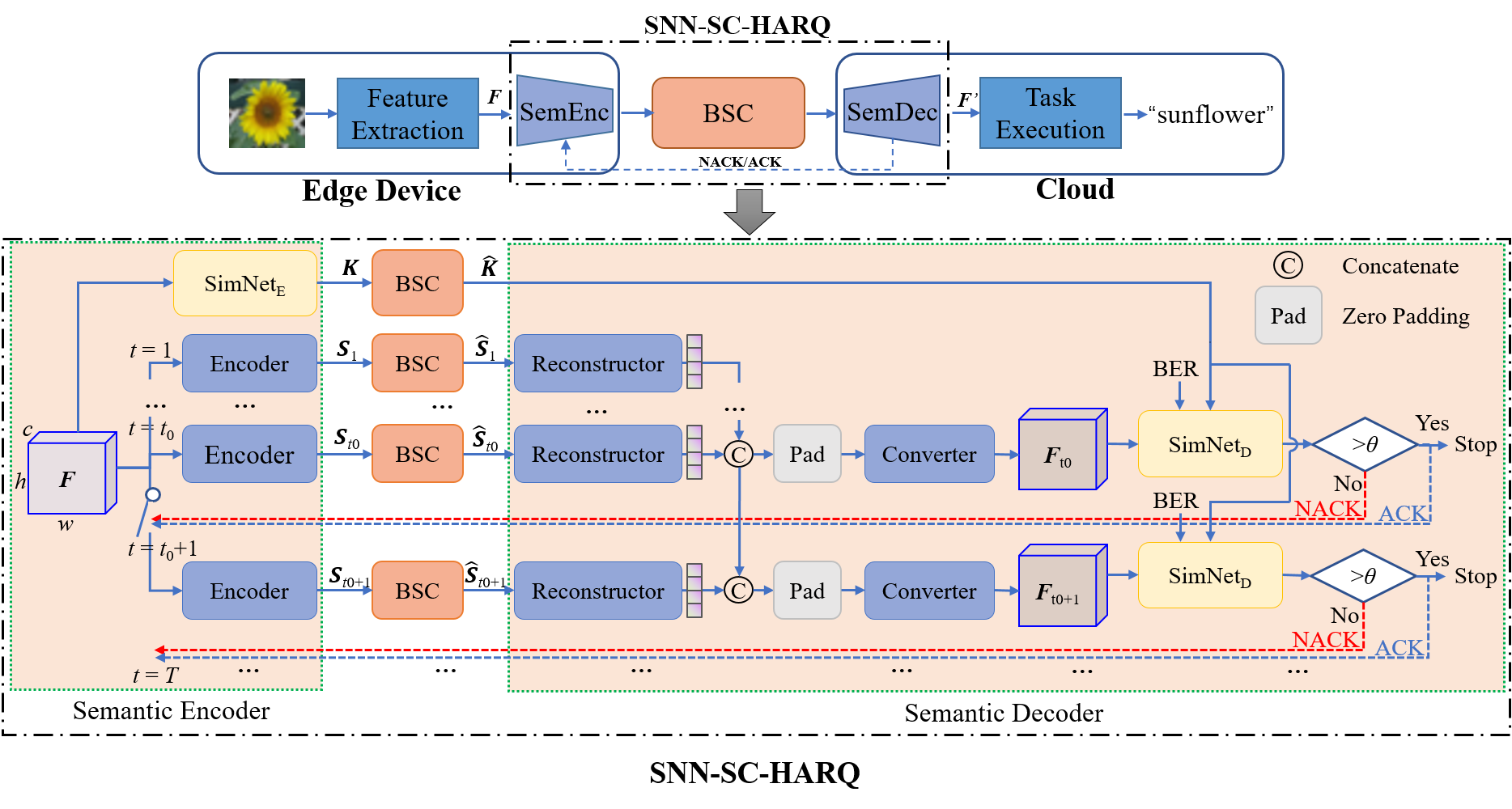}
  \caption{The structure of the SNN-SC-HARQ. The encoder, reconstructor, and converter make up the multi-rate SNN-SC. The SimNet$_\text{E}$ and SimNet$_\text{D}$ make up the SimNet.}
  \label{fig:1}
\end{figure*}
\section{Methodology}
\label{sec:2}
\subsection{System Model}
As illustrated in Fig. \ref{fig:1}, an image classification model is divided into the feature extraction part on the edge device and the task execution part on the cloud. To compress and transmit features over BSC, SNN-SC-HARQ is utilized. SNN-SC-HARQ is comprised of two components: a semantic similarity network SimNet and a feature SC model multi-rate SNN-SC. SimNet is composed of SimNet$_\text{E}$ and SimNet$_\text{D}$, and multi-rate SNN-SC is composed of an encoder, reconstructor, and converter.

For the feature $\boldsymbol{F}$, SimNet$_\text{E}$ firstly extracts binary prior information $\boldsymbol{K}$ from $\boldsymbol{F}$ then sends it to the cloud through BSC. The disturbed prior information $\boldsymbol{\hat{K}}$ is received in the cloud. Subsequently, the multi-rate SNN-SC transmits $\boldsymbol{F}$ in an incremental bandwidth manner. The transmission process of multi-rate SNN-SC is divided into two stages: initial transmission and retransmission stage. In the initial transmission stage, the SNN-based encoder and the reconstructor operate for $t_0$ time steps. At time step $t$, the encoder extracts and transmits the compact semantic information $\boldsymbol{S}_t$, which is then recovered by the reconstructor. The outputs of the reconstructor over $t_0$ time steps are then concatenated, zero-padded, and summed by the FCN-based converter to get the initial reconstruction result, $\boldsymbol{F}_{t0}$. The semantic similarity between $\boldsymbol{F}$ and $\boldsymbol{F}_{t0}$ is calculated by SimNet$_\text{D}$, taking into account the reconstruction result $\boldsymbol{F}_{t0}$, $\boldsymbol{\hat{K}}$, and the Bit Error Rate (BER) of BSC. If the semantic similarity exceeds a threshold $\theta$, the cloud will feedback ACK, and the $\boldsymbol{F}_{t0}$ is designated as the final reconstructed feature $\boldsymbol{F'}$. Otherwise, the cloud will feedback NACK, and retransmission will be initiated. In the retransmission stage, the encoder and reconstructor of multi-rate SNN-SC will continue to run another time step to transmit incremental information and repeat the reconstruction process until semantic similarity exceeds the threshold or the maximum running time step $T$ is reached. Finally, the reconstructed feature $\boldsymbol{F'}$ is fed into the task execution module to obtain the classification result.

\subsection{Multi-Rate SNN-SC}
\begin{figure}
  \centering
  \includegraphics[width=0.75\columnwidth]{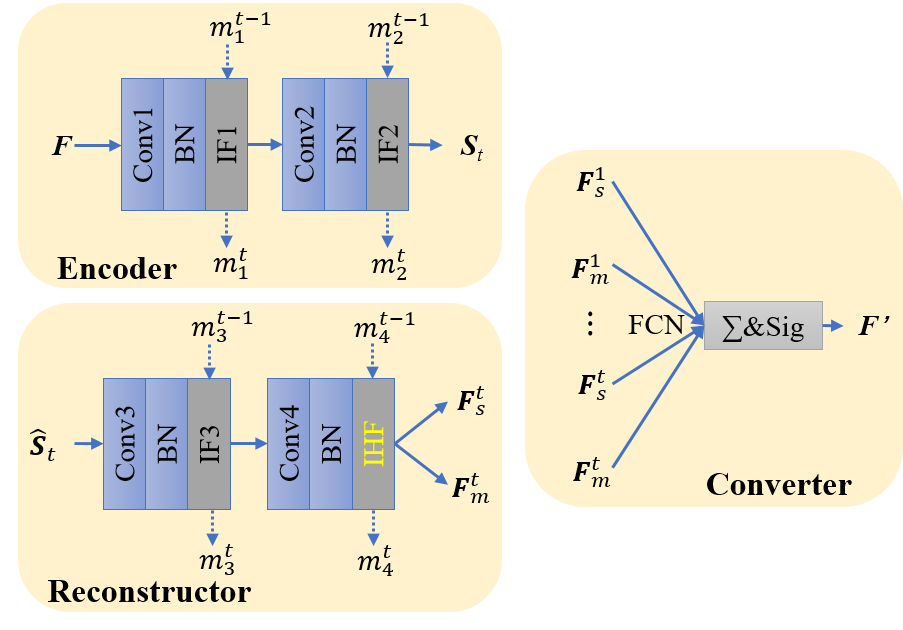}
  \caption{The detailed structure of the multi-rate SNN-SC}
  \label{fig:7}
\end{figure}
The detailed structure of the encoder, reconstructor, and converter of multi-rate SNN-SC is shown in Fig. \ref{fig:7}. During the running of multi-rate SNN-SC, all parameters are shared across all time steps. The spiking neurons utilized in this model are the Integrate-Fire (IF) \cite{12} and the Integrate and Hybrid Fire (IHF) \cite{6}. At each time step, IF executes three serial steps: charge, fire, and reset \cite{34}, which can be described as follows:
\begin{equation}
m_{i}^{t}=  m_{i}^{t-1} + I_{i-1}^{t}, \label{equ:2}
\end{equation}
\begin{equation}
S_{i}^{t}= 
\begin{cases}
1,\quad m_{i}^{t}>V_\text{th},\\
0,\quad \text{otherwise}, \label{equ:3}
\end{cases}
\end{equation}
\begin{equation}
m_{i}^{t} = 
\begin{cases}
(1-S_{i}^{t})m_{i}^{t}+S_{i}^{t}V_\text{reset}\quad &\text{(hard reset)}, \\
m_{i}^{t}-S_{i}^{t}V_\text{th} \quad &\text{(soft reset)}, \label{equ:4}
\end{cases}
\end{equation}
where $m_{i}^{t}$ represents the membrane potential of $i$th neuron, $I_{i-1}^{t}$ is the input from previous layer, and $S_{i}^{t}$ is the output spike at time step $t$. $V_\text{th}$ in Eq. (\ref{equ:3}) is the firing threshold of the membrane potential, and $V_\text{reset}$ in Eq. (\ref{equ:4}) is the reset value of membrane potential. Eq. (\ref{equ:2}) is the charging process, where IF accumulates the potential of the previous time step and the current input as the current potential. After charging, the potential is compared to the threshold $V_\text{th}$ to determine whether to fire spikes, as shown in Eq. (\ref{equ:3}). After firing, the potential is reset in Eq. (\ref{equ:4}). The IHF and IF operation procedures are similar, except that IHF additionally outputs the reset potential value $m_{i}^{t}$. With IF, encoder extracts semantic information $\boldsymbol{S}_t$ from $\boldsymbol{F}$. With IF and IHF, the reconstructor decodes the disturbed semantic information $\boldsymbol{\hat{S}}_t$ and outputs two results: $\boldsymbol{F}^t_s$ and $\boldsymbol{F}^t_m$, where $\boldsymbol{F}^t_s$ contains spike information and $\boldsymbol{F}^t_m$ contains membrane potential information. Finally, the converter uses FCN to summarize the outputs of reconstructor to get the final reconstructed feature $\boldsymbol{F'}$. 

To make multi-rate SNN-SC adapt to different running time steps, we use different running time steps to train it. If the selected time step is less than the maximum preset time step, the reconstructor's outputs will be zero-padded and then inputted into the converter. The training method is described in Algorithm \ref{alg:2}. $f_\text{E}(\cdot;\boldsymbol{\mu})$ and $f_\text{T}(\cdot;\boldsymbol{\lambda})$ represent the feature extraction and task execution parts of the image classification model with parameters $\boldsymbol{\mu}$ and $\boldsymbol{\lambda}$, $E(\cdot;\boldsymbol{\alpha})$, $R(\cdot;\boldsymbol{\beta})$, and $C(\cdot;\boldsymbol{\gamma})$ is the encoder, reconstructor, and converter with parameters $\boldsymbol{\alpha}$, $\boldsymbol{\beta}$, and $\boldsymbol{\gamma}$. The BSC is modeled as $B(\cdot;p)$ with BER $p$. 

\begin{algorithm}
	\caption{Training Algorithm for Multi-Rate SNN-SC} 
	\label{alg:2} 
	\begin{algorithmic}
		\renewcommand{\algorithmicrequire}{\textbf{Input:} }
		\renewcommand{\algorithmicensure}{\textbf{Output:}}
		\Require{Image $\boldsymbol{X}$, classification label $\boldsymbol{L}$}
		\Ensure{Multi-Rate SNN-SC model parameters $(\boldsymbol{\alpha}, \boldsymbol{\beta}, \boldsymbol{\gamma})$}
		\State $\boldsymbol{F} = f_\text{E}(\boldsymbol{X};\boldsymbol{\mu})$     \Comment{Extract feature on the device}
		\State Randomly sample time step $t$ from [$t_0, t_0+1, ..., T$]
		\State Randomly sample BER $p$
		\State Initialize Outputs = []
		\For{i = 1 to $t$} 
		\State $\boldsymbol{S}_i = E(\boldsymbol{F};\boldsymbol{\alpha}) $  \Comment{Encoder}
		\State $\boldsymbol{\hat{S}}_i = B(\boldsymbol{S}_i;p)$ \Comment{Transmit over BSC}
		\State $\boldsymbol{F}^i_s, \boldsymbol{F}^i_m = R(\boldsymbol{\hat{S}}_i; \boldsymbol{\beta})$ \Comment{Reconstructor}
		\State Outputs = Concate(Outputs, $\boldsymbol{F}^i_s, \boldsymbol{F}^i_m$) \Comment{Concate output}
		\EndFor
		\State Padding Outputs with zeros
		\State $\boldsymbol{F'} = C(\text{Outputs};\boldsymbol{\gamma})$  \Comment{Converter}
		\State $\boldsymbol{P} = f_\text{T}(\boldsymbol{F'}; \boldsymbol{\lambda})$   \Comment{Execute task on the cloud}
		\State Compute loss with loss function $\mathcal L$
		\State Update model parameters $(\boldsymbol{\alpha}, \boldsymbol{\beta}, \boldsymbol{\gamma})$ by gradient descent
	\end{algorithmic} 

\end{algorithm}
During the training, the running time step is randomly sampled from [$t_0, t_0+1,..., T$] (1$\leq t_0 <T$ ), where $t_0$ and $T$ are the minimum and maximum time step respectively. The loss function is defined as follows and has been proven effective in \cite{6}:
\begin{equation}
\label{equ:1}
    \mathcal L =  \mathcal L_{\text{CE}}( \boldsymbol{P}, \boldsymbol{L}) + (\frac{1}{t} \sum_{i =1}^{t}H(\boldsymbol{S}_i)-1)^2,
\end{equation}
where $\mathcal L_{\text{CE}}$ represents the cross-entropy between the predicted classification result $\boldsymbol{P}$ and the label $\boldsymbol{L}$, and $H(\boldsymbol{S}_i)$ represents the entropy of extracted semantic information $\boldsymbol{S}_i$. As $\boldsymbol{S}_i$ is exclusively composed of 0s and 1s, its entropy can be estimated by calculating the frequency of these values. To compute the gradient of spiking neurons, we follow the surrogate gradient approach \cite{35} by replacing the firing step with a differentiable surrogate function, namely a sigmoid $\sigma(x) = (1 + e^{-x})^{-1}$.

After training, multi-rate SNN-SC can reconstruct features under different running time steps with a single model. In other words, multi-rate SNN-SC is capable of transmitting features at different bandwidths. Based on trained multi-rate SNN-SC, a policy network is proposed to choose an bandwidth.

\subsection{Feature Semantic Similarity Network}
In traditional HARQ-based communication systems, cyclic redundancy check (CRC) code is used to detect errors at the symbol level. However, the CRC may classify a reconstructed feature as erroneous even if it achieves satisfactory task performance but differs from the original feature. Therefore, we propose a policy network to replace the CRC module.

We argue that if the reconstructed features maintain high task performance in the cloud, the inference results of the reconstructed features on the task execution module will be similar to that of the original features. The similarity between the original feature $\boldsymbol{F}$ and $\boldsymbol{F'}$ can be measured as
\begin{equation}
\label{eq:2}
Sim( \boldsymbol{F}, \boldsymbol{F'}) = \frac{f_\text{T}(\boldsymbol{F};\boldsymbol{\lambda})f_\text{T}(\boldsymbol{F'}; \boldsymbol{\lambda})^\mathrm{T}}{|f_\text{T}(\boldsymbol{F}; \boldsymbol{\lambda})||f_\text{T}(\boldsymbol{F'}; \boldsymbol{\lambda})|},
\end{equation}
where we input $\boldsymbol{F}$ and $\boldsymbol{F'}$ into task execution module $f_\text{T}(\cdot;\boldsymbol{\lambda})$, and then calculate the cosine similarity of the inference results, which is referred to as the feature semantic similarity. The higher the semantic similarity, the more task-related information is preserved in the reconstructed features.
\begin{figure}
  \centering
  \includegraphics[width=0.9\columnwidth]{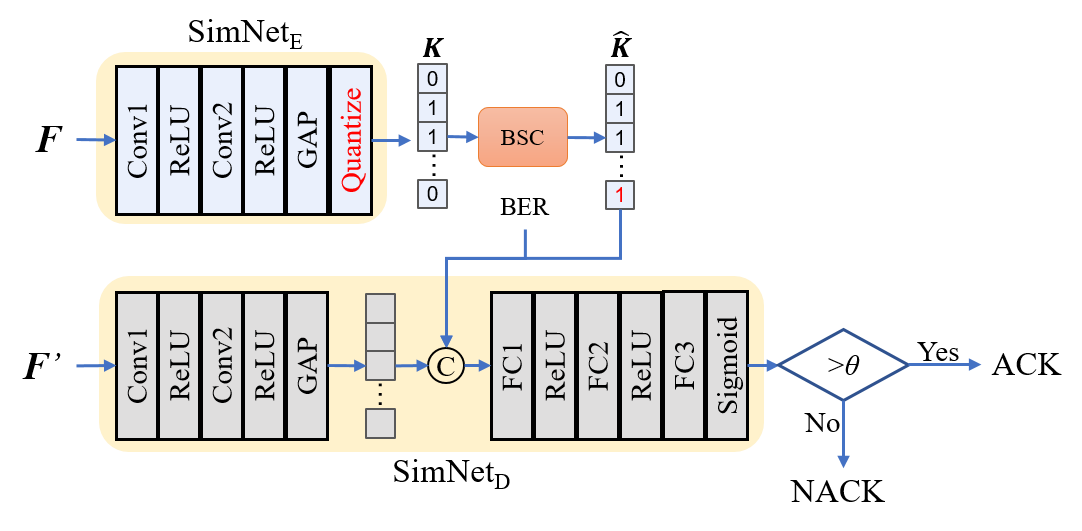}
  \caption{The structure of the feature semantic similarity network SimNet}
  \label{fig:2}
\end{figure}
However, the original feature $\boldsymbol{F}$ is not accessible in the cloud, which renders it impossible for the receiver to directly compute semantic similarity. Consequently, we propose SimNet to estimate the feature semantic similarity between $\boldsymbol{F}$ and $\boldsymbol{F'}$. The detailed structure of SimNet is shown in Fig. \ref{fig:2}. On the edge device, the prior information $\boldsymbol{K}$ is extracted by SimNet${_\text{E}}$ and transmitted over BSC. To make $\boldsymbol{K}$ compatible with BSC, 1-bit quantization is used in SimNet$_\text{E}$. On the cloud, SimNet$_\text{D}$ utilizes $\boldsymbol{F'}$, noisy prior information $\boldsymbol{\hat{K}}$, and the BER $p$ of BSC to estimate the feature semantic similarity. Based on the trained multi-rate SNN-SC, SimNet is trained using the actual semantic similarity between $\boldsymbol{F}$ and $\boldsymbol{F'}$ as labels. Denote SimNet$_\text{E}$ and SimNet$_\text{D}$ as $S_\text{E}(\cdot;\boldsymbol{\omega})$ and $S_\text{D}(\cdot;\boldsymbol{\phi})$ with parameters $\boldsymbol{\omega}$ and $\boldsymbol{\phi}$, the training method is shown in Algorithm \ref{alg:3}.

\begin{algorithm} 
	\caption{Training Algorithm for SimNet} 
	\label{alg:3} 
	\begin{algorithmic}
		\renewcommand{\algorithmicrequire}{\textbf{Input:} }
		\renewcommand{\algorithmicensure}{\textbf{Output:}}
		\Require{Image $\boldsymbol{X}$}
		\Ensure{SimNet model parameters $(\boldsymbol{\omega}, \boldsymbol{\phi})$}
		\State $\boldsymbol{F} = f_\text{E}(\mathbf{X};\boldsymbol{\mu})$     \Comment{Extract feature on the device}
		\State  $\boldsymbol{K} = S_\text{E}(\mathbf{F};\boldsymbol{\omega})$    \Comment{Extract prior information}
		\State Randomly sample time step $t$ from [$t_0, t_0+1, ..., T$]
		\State Randomly sample BER $p$
		\State  $\boldsymbol{\hat{K}} = B(\boldsymbol{K};p)$   \Comment{Transmit over BSC}
		\State Generate $\boldsymbol{F'}$ by multi-rate SNN-SC with $p$ and $t$
		\State Compute label $Sim(\boldsymbol{F},\boldsymbol{F'})$ by Eq. (\ref{eq:2})
		\State Compute loss $\mathcal L = (S_\text{D}(\boldsymbol{F'}, \boldsymbol{\hat{K}}, p; \boldsymbol{\phi}) - Sim(\boldsymbol{F},\boldsymbol{F'}))^2$
		\State Update model parameters $\boldsymbol{\omega}, \boldsymbol{\phi}$ by gradient descent
	\end{algorithmic} 
\end{algorithm}

With SimNet, we can get semantic similarity between reconstructed and original features in the cloud. By comparing the feature semantic similarity with a threshold $\theta$, SimNet can make decisions at the semantic level, overcoming the disadvantage of CRC. Finally, the SNN-SC-HARQ is obtained by integrating multi-rate SNN-SC with SimNet.

\section{Experiments}
\label{sec:3}
\subsection{Experiment Settings}
\label{sec:4-1}
The experiments are conducted on CIFAR-100, and the image classification model used in the CI scenario is ResNet50 \cite{14}. The ResNet50 and SimNet are implemented with PyTorch, and the multi-rate SNN-SC is developed using SpikingJelly \cite{16}. Experiments are conducted on Tesla V100.

The trained ResNet50 model obtains a classification accuracy of 77.81\% on the CIFAR-100, and it is divided into feature extraction part and task execution part at the output of its 14th bottleneck block, where the feature dimension is (2048,4,4), corresponding to (channels, height, width). Then the feature is transmitted by SNN-SC-HARQ over BSC. The classification accuracy of the cloud task execution part serves as the evaluation metric of the SNN-SC-HARQ.

To get a well-trained model, a multi-step training strategy is employed. Initially, the parameters of the trained ResNet50 are kept fixed, and the multi-rate SNN-SC model is trained using Algorithm \ref{alg:2}. Subsequently, ResNet50 and the multi-rate SNN-SC are fine-tuned jointly. Finally, SimNet is trained using Algorithm \ref{alg:3}, with the parameters of ResNet50 and multi-rate SNN-SC being fixed. During training, each transmission randomly chooses a channel BER from [0, 0.3].

In SNN-SC-HARQ, the output dimension of SimNet$_\text{E}$ is (32,1,1), indicating that 32 bits will be transmitted at the very beginning. The output dimension of the multi-rate SNN-SC encoder is (32,4,4), meaning that 512 bits will be transmitted per time step. The minimum and maximum time steps, denoted as $t_0$ and $T$ respectively, for running the multi-rate SNN-SC are set to 4 and 8. Therefore, the SNN-SC-HARQ supports five possible bandwidths: 2080 bits (4 $\times$ 512 + 32 = 2080), 2592 bits, 3104 bits, 3616 bits, and 4128 bits.

\subsection{Effectiveness of SimNet}
\label{sec:4-2}
\begin{figure}
  \centering
  \includegraphics[width=\columnwidth]{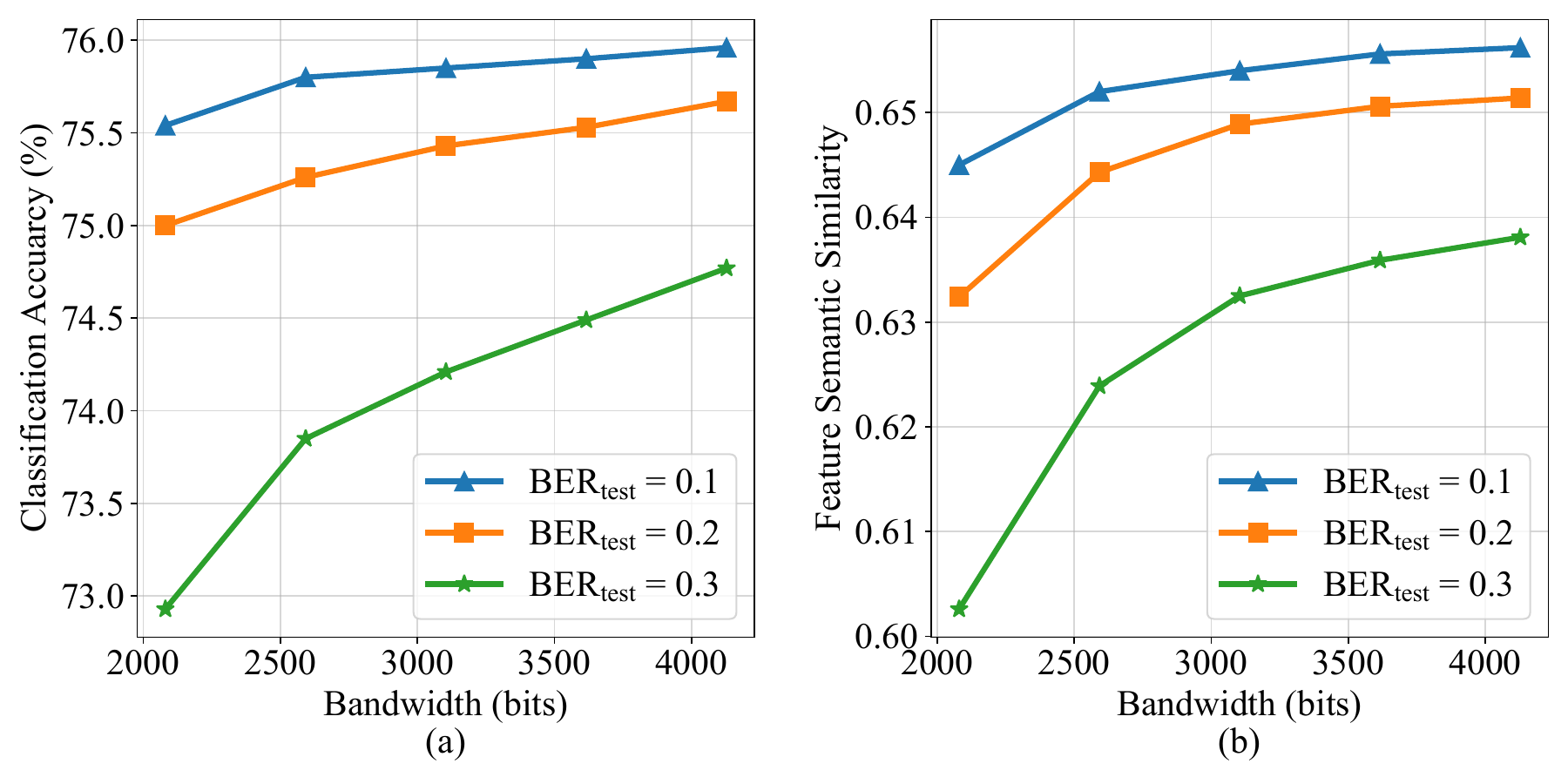}
  \caption{Classification accuracy and feature semantic similarity under different bandwidths and BERs.}
  \label{fig:3}
\end{figure}

To validate the efficacy of SimNet, the correlation between semantic similarity and the performance of classification tasks on the cloud is analyzed. The multi-rate SNN-SC is evaluated across various bandwidths and BERs, the semantic similarity estimated by SimNet and the classification accuracy caused by the reconstructed features are shown in Fig. \ref{fig:3}.

%\begin{figure*}[htbp]
%	\centering
%	\subfloat[]{\includegraphics[width=0.57\columnwidth]{BER_Bits_Acc.pdf}}
%	\subfloat[]{\includegraphics[width=0.5\columnwidth]{BER_Acc.pdf}}
%	\subfloat[]{\includegraphics[width=0.5\columnwidth]{BER_Bits.pdf}}
%	\caption{The average bandwidth and average classification accuracy of the SNN-SC and SNN-SC-HARQ for different BERs and thresholds.}
%	\label{fig:4}
%\end{figure*}

Figures Fig. \ref{fig:3}(a) and Fig. \ref{fig:3}(b) each contain three curves representing classification accuracy and feature semantic similarity under different bandwidths, with channel BERs of 0.1, 0.2, and 0.3 (i.e., BER${_\mathrm{test}}$). In Fig. \ref{fig:3}(a), classification accuracy decreases with increasing BER under the same bandwidth, due to noise in the BSC. Conversely, classification accuracy increases with bandwidth when BER is fixed, as more data is received by the cloud. In Fig. \ref{fig:3}(b), semantic similarity exhibits a similar trend to classification accuracy when bandwidth or BER changes. The Pearson correlation coefficients of semantic similarity and classification curve are 0.995, 0.958, and 0.987 when BER${_\mathrm{test}}$ is 0.1, 0.2, and 0.3, respectively. Therefore, we can conclude that semantic similarity is positively correlated with classification accuracy.

By comparing the semantic similarity of the features with a preset threshold, an error detection function similar to CRC is implemented. When the semantic similarity of the features exceeds the threshold, we predict that the reconstructed features will achieve satisfactory classification performance when fed into the task execution module. Subsequently, an ACK signal is feedback to the edge device to halt the transmission of data by the multi-rate SNN-SC. Conversely, a NACK signal is feedback to the edge device, prompting the multi-rate SNN-SC to increase bandwidth by running another time step and repeating the process until the similarity exceeds the threshold or the multi-rate SNN-SC reaches its maximum time step. In summary, SimNet can serve as a policy module based on semantic similarity and is more suitable for semantic communication than CRC.
%Although SimNet adds an extra 32 bits of bandwidth, its impact on the overall bandwidth is negligible compared to the multi-rate SNN-SC that transmits 512 bits per time step.

\begin{figure}
	\centering
	\includegraphics[width=0.8\columnwidth]{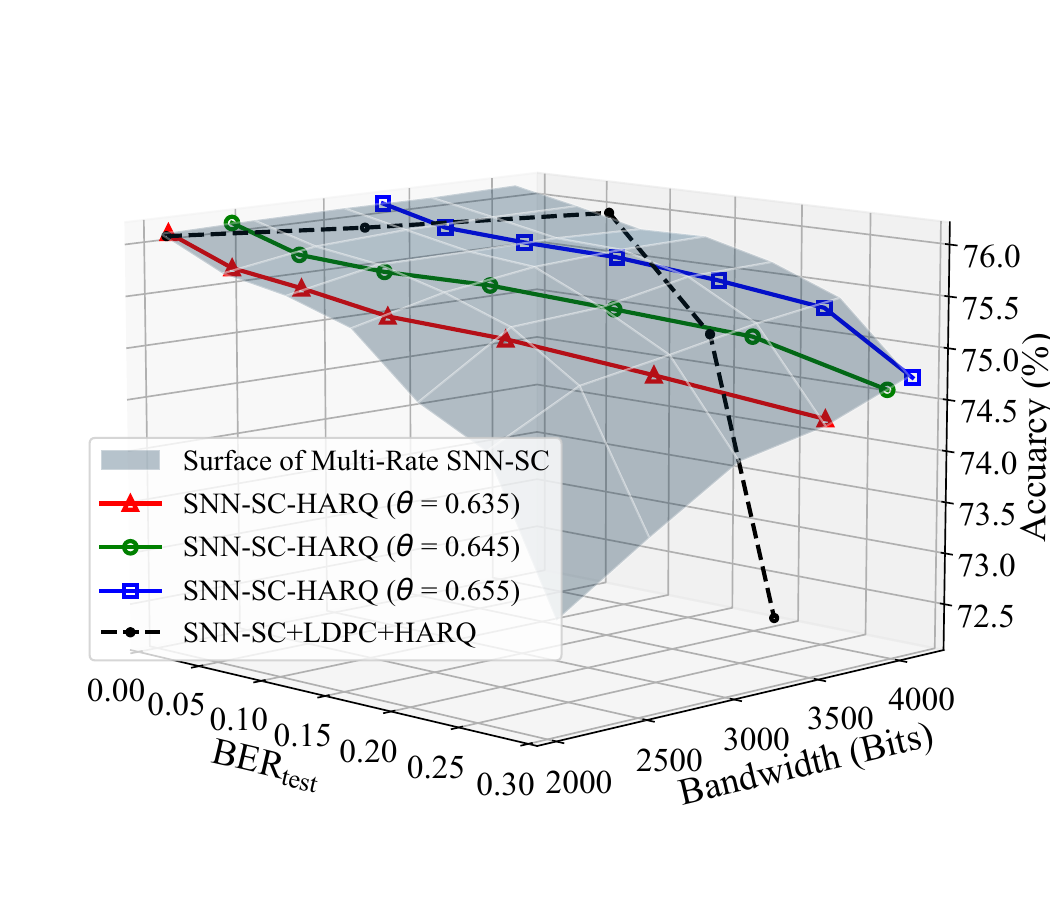}
	\caption{The bandwidth and classification accuracy of the SNN-SC-HARQ, Multi-Rate SNN-SC and SNN-SC+LDPC+HARQ}
	\label{fig:4}
\end{figure}

\subsection{Performance of SNN-SC-HARQ}
In this section, we utilize the semantic communication method Multi-Rate SNN-SC and the separate coding method SNN-SC+LDPC+HARQ as our baseline methods. For Multi-Rate SNN-SC, we manually adjust its transmission bandwidth and evaluate its performance under various BERs to obtain a performance surface. As for SNN-SC+LDPC+HARQ, we employ SNN-SC\cite{6} as the source encoder. To ensure that the amount of transmitted data matches that of SNN-SC-HARQ, the transmission bandwidth of SNN-SC is fixed at 1536 bits ($t$ = 3) and trained under noise-free conditions. The encoding results are then encoded to 4128 bits using LDPC. Initially, the transmission bandwidth is set at 2080 bits, and in the case of CRC error detection, 512 bits are retransmitted.

\begin{table}[htbp]
\caption{Absolute Performance gap between SNN-SC-HARQ and Multi-rate SNN-SC under the same bandwidth}
\centering
\begin{tabular}[width=1.5\columnwidth]{c||c|c|c|c|c|c|c}
\hline
\hline
\diagbox{$\theta$}{BER}&0.00&0.05 &0.10&0.15 &0.20&0.25 &0.30 \\ 
\hline
\hline
0.635 &  0.02 & 0.06 & 0.09 & 0.01 & 0.03 & 0.11 & 0.07 \\ 

0.645 &  0.02 & 0.01 & 0.07 & 0.09 & 0.06 & 0.11 & 0.01 \\ 

0.655 &  0.00 & 0.03 & 0.01 & 0.06 & 0.08 & 0.05 & 0.05 \\ 
\hline
\end{tabular}
\label{tab:1}
\end{table}

Figure \ref{fig:4} depicts the changes in classification accuracy and transmission bandwidth as a function of channel bit error rate (BER) for both SNN-SC-HARQ and baseline methods. The thresholds for SNN-SC-HARQ are set to 0.635, 0.645, and 0.655, respectively. Regarding classification accuracy, SNN-SC+LDPC+HARQ achieves optimal performance when the BER is less than 0.1, as LDPC can correct all transmission errors. However, when the BER exceeds 0.1, LDPC fails to work normally, resulting in a sharp decline in classification performance, known as the `cliff effect'. In contrast, SNN-SC-HARQ and Multi-Rate SNN-SC exhibit graceful degradation of performance under poor channel conditions, highlighting the superiority of SC. In terms of transmission bandwidth, SNN-SC+LDPC+HARQ can automatically adjust its bandwidth. However, the transmission bandwidth experiences a sharp increase when the BER exceeds 0.05, ultimately reaching its upper limit of transmission bandwidth when the BER reaches 0.15. SNN-SC-HARQ gradually increases the transmission bandwidth as the channel quality drops, with a larger SNN-SC-HARQ threshold leading to a bigger transmission bandwidth. Multi-Rate SNN-SC lacks the capability to automatically adjust bandwidth and requires manual configuration, which is inconvenient. Additionally, we use the interpolation algorithm to determine the classification performance of Multi-Rate SNN-SC on the performance surface when the bandwidth is equivalent to that of SNN-SC-HARQ. Table \ref{tab:1} illustrates the absolute performance gap between Multi-Rate SNN-SC and SNN-SC-HARQ. The values in Table 1 are all very close to 0, suggesting that the proposed SNN-SC-HARQ achieves adaptive bandwidth adjustment without losing classification performance compared to Multi-Rate SNN-SC.

\section{Conclusion}
\label{sec:4}
Recently, SNN-based SC models have demonstrated their effectiveness in transmitting features over BSC. However, the bandwidths of these models are fixed and cannot be adjusted adaptively. In this paper, we propose a novel SNN-based feature SC model, SNN-SC-HARQ, to address this issue. SNN-SC-HARQ consists of two models: multi-rate SNN-SC and SimNet. Multi-rate SNN-SC can reconstruct features under different bandwidths within a single model. SimNet is built based on multi-rate SNN-SC and can estimate the semantic similarity between the features reconstructed by multi-rate SNN-SC and the original features. By comparing the semantic similarity with a preset threshold, SimNet performs a semantic level check. Experimental results show that SNN-SC-HARQ not only overcomes the cliff effect, but also realizes adaptive bandwidth adjustment.

\bibliographystyle{IEEEtran}
\bibliography{refs}

\begin{thebibliography}{10}
\providecommand{\url}[1]{#1}
\csname url@samestyle\endcsname
\providecommand{\newblock}{\relax}
\providecommand{\bibinfo}[2]{#2}
\providecommand{\BIBentrySTDinterwordspacing}{\spaceskip=0pt\relax}
\providecommand{\BIBentryALTinterwordstretchfactor}{4}
\providecommand{\BIBentryALTinterwordspacing}{\spaceskip=\fontdimen2\font plus
\BIBentryALTinterwordstretchfactor\fontdimen3\font minus
  \fontdimen4\font\relax}
\providecommand{\BIBforeignlanguage}[2]{{%
\expandafter\ifx\csname l@#1\endcsname\relax
\typeout{** WARNING: IEEEtran.bst: No hyphenation pattern has been}%
\typeout{** loaded for the language `#1'. Using the pattern for}%
\typeout{** the default language instead.}%
\else
\language=\csname l@#1\endcsname
\fi
#2}}
\providecommand{\BIBdecl}{\relax}
\BIBdecl

\bibitem{2}
Y.~Kang, J.~Hauswald, C.~Gao, A.~Rovinski, T.~Mudge, J.~Mars, and L.~Tang,
  ``Neurosurgeon: Collaborative intelligence between the cloud and mobile
  edge,'' \emph{SIGARCH Comput. Archit. News}, vol.~45, no.~1, pp. 615--629,
  2017.

\bibitem{28}
Z.~Qin, X.~Tao, J.~Lu, W.~Tong, and G.~Y. Li, ``Semantic communications:
  Principles and challenges,'' \emph{arXiv preprint arXiv:2201.01389}, 2021.

\bibitem{3}
M.~Jankowski, D.~G{\"u}nd{\"u}z, and K.~Mikolajczyk, ``Wireless image retrieval
  at the edge,'' \emph{IEEE J. Sel. Areas Commun.}, vol.~39, no.~1, pp.
  89--100, 2020.

\bibitem{4}
J.~Shao and J.~Zhang, ``Bottlenet++: An end-to-end approach for feature
  compression in device-edge co-inference systems,'' in \emph{Proc. IEEE Int.
  Conf. Commun. Workshop}, 2020, pp. 1--6.

\bibitem{5}
M.~Wang, Z.~Zhang, J.~Li, M.~Ma, and X.~Fan, ``Deep joint source-channel coding
  for multi-task network,'' \emph{IEEE Signal Process. Lett.}, vol.~28, pp.
  1973--1977, 2021.

\bibitem{6}
M.~Wang, J.~Li, M.~Ma, and X.~Fan, ``S-jscc: A digital joint source-channel
  coding framework based on spiking neural network,'' \emph{arXiv preprint
  arXiv:2210.06836}, 2022.

\bibitem{7}
W.~Li, Z.~Ma, L.-J. Deng, X.~Fan, and Y.~Tian, ``Neuron-based spiking
  transmission and reasoning network for robust image-text retrieval,''
  \emph{IEEE Trans. Circuits Syst. Video Technol.}, 2022.

\bibitem{10}
P.~Jiang, C.-K. Wen, S.~Jin, and G.~Y. Li, ``Deep source-channel coding for
  sentence semantic transmission with harq,'' \emph{IEEE Trans. Commun.},
  vol.~70, no.~8, pp. 5225--5240, 2022.

\bibitem{11}
Q.~Zhou, R.~Li, Z.~Zhao, Y.~Xiao, and H.~Zhang, ``Adaptive bit rate control in
  semantic communication with incremental knowledge-based harq,'' \emph{IEEE
  Open J. Commun. Soc.}, vol.~3, pp. 1076--1089, 2022.

\bibitem{12}
S.~Lu and A.~Sengupta, ``Exploring the connection between binary and spiking
  neural networks,'' \emph{Frontiers Neurosci.}, vol.~14, p. 535, 2020.

\bibitem{34}
W.~Fang, Z.~Yu, Y.~Chen, T.~Masquelier, T.~Huang, and Y.~Tian, ``Incorporating
  learnable membrane time constant to enhance learning of spiking neural
  networks,'' in \emph{Proc. IEEE/CVF Conf Comput. Vis. Pattern Recognit.
  (CVPR)}, 2021, pp. 2661--2671.

\bibitem{35}
E.~O. Neftci, H.~Mostafa, and F.~Zenke, ``Surrogate gradient learning in
  spiking neural networks: Bringing the power of gradient-based optimization to
  spiking neural networks,'' \emph{IEEE Signal Process. Mag.}, vol.~36, no.~6,
  pp. 51--63, 2019.

\bibitem{14}
K.~He, X.~Zhang, S.~Ren, and J.~Sun, ``Deep residual learning for image
  recognition,'' in \emph{Proc. IEEE/CVF Conf Comput. Vis. Pattern Recognit.
  (CVPR)}, 2016, pp. 770--778.

\bibitem{16}
W.~Fang, Y.~Chen, J.~Ding, D.~Chen, Z.~Yu, H.~Zhou, Y.~Tian, and other
  contributors, ``Spikingjelly,''
  \url{https://github.com/fangwei123456/spikingjelly}, 2020, accessed:
  2021-07-03 (0.0.0.0.6).

\end{thebibliography}
\end{document}